\documentclass[conference]{IEEEtran}

\usepackage{graphicx}
\usepackage{amsmath}
\usepackage{amsfonts}
\usepackage{amssymb}
\usepackage{subfigure}
\usepackage{pgfplots}

\newtheorem{definition}{Definition}

\newtheorem{condition}{Condition}

\newtheorem{example}{Example}

\begin{document}

\title{\huge{Lattice Partition Multiple Access: A New Method of Downlink Non-orthogonal Multiuser Transmissions}}

\author{\IEEEauthorblockN{Dong Fang${}^\dag$, Yu-Chih Huang${}^\ddag$, Zhiguo Ding${}^\natural$, Giovanni Geraci${}^\dag$, Shin-Lin Shieh${}^\ddag$, and Holger Claussen${}^\dag$}
\IEEEauthorblockA{${}^\dag$Department of Small Cells Research, Bell Labs, Nokia, Ireland\\} \IEEEauthorblockA{${}^\ddag$Department of Communication Engineering, National Taipei University, Taipei, Taiwan\\} \IEEEauthorblockA{${}^\natural$School of Computing and Communications, Lancaster University, UK\\}
Emails: ${}^\dag$\{\texttt{dong.fang, giovanni.geraci, holger.claussen}\}\texttt{@nokia.com}\\
${}^\ddag$\{\texttt{ychuang, slshieh}\}\texttt{@mail.ntpu.edu.tw}, ${}^\natural$\texttt{z.ding}\texttt{@lancaster.ac.uk}}


\maketitle

\begin{abstract}
In this paper, we propose a new downlink non-orthogonal multiuser superposition transmission scheme for future 5G cellular networks, which we refer to as the lattice partition multiple access (LPMA). In this proposed design, the base station transmits multilevel lattice codes for multiple users. Each user's code level corresponds to a distinct prime and is weighted by a product of all distinct primes of the other users excluding its own. Due to the structural property of lattice codes, each user can cancel out the interference from the other code levels by using the modulo lattice operation in a successive/parrallel manner. LPMA can overcome the drawback of non-orthogonal multiple access (NOMA), which arises when users have similar channel conditions. We demonstrate that the proposed LPMA shows a clear throughput enhancement over the current NOMA scheme.
\end{abstract}


\IEEEpeerreviewmaketitle

\section{Introduction}
The International Mobile Telecommunications (IMT)-2020 promotion group drafted a white paper of `the vision and demand of 5G' \cite{IMT_2020} in 2014 to specify the key performance indicators of the fifth generation (5G) cellular networks, i.e., the 1000 times future traffic growth, the 100 times connectivity requirements, and the 5 to 15 times spectrum efficiency enhancement etc.. To this end, new multiple access (MA) techniques must be developed because current 4G MA solutions have not been designed to meet such demands.
As a promising multiple access technique, considerable attention has been drawn to the non-orthogonal multiple access (NOMA) \cite{Saito} due to its less processing overhead and superior spectral efficiency. The 3rd Generation Partnership Project (3GPP) initiated several discussions about the benefits of non-orthogonal downlink transmissions. In these meetings, NOMA has  been selected as the study item in Long-term Evolution (LTE) release 13, termed multiuser superposition transmission (MUST) \cite{3GPP_SI_TR}. NOMA multiplexes users on the power-domain at the transmitter side where far users are allocated more transmission power while near users are allocated less transmission power. As such, far users decode their desired signals by treating others' information as noise. On the other hand, near users adopt the successive interference cancellation (SIC) to extract their desired signals, i.e., they first decode the signals of far users and then decode their own after subtracting the decoded signals of far users.

However, in current NOMA schemes, it is not specified which coding scheme is used. Moreover, the scenarios where the users have similar channel conditions could diminish the performance gain of NOMA over orthogonal multiple access (OMA), such as TDMA, FDMA etc.. To address this challenge, usually,  the geometrical separation should be fulfilled by the scheduler which only clusters the users with different channel gains for superposition transmissions. Such a scheduler results in an excessively high complexity if the cellular network has a high user density and huge traffic demand. In addition, in order to achieve geometrical separation, the scheduler could hardly project the beams onto user clusters with maximal spatial orthogonality so that the crosstalk occurs among the neighbouring user clusters, which is another challenge especially for a MIMO-NOMA \cite{MIMO_NOMA} system.

Motivated by the above challenges of NOMA, the objective of this paper is to provide a new downlink superposition transmission scheme so that the drawback of NOMA which arises when users have similar channel conditions can be overcome. This new superposition transmission scheme is based on multilevel lattice codes, referred to as lattice partition multiple access (LPMA). LPMA exploits the structural property of the lattice codes to harness the co-channel interference among users. In the scenario where users have the similar channel conditions, LPMA can multiplex users by assigning them different lattice partitions which are isomorphic to the same finite field. We demonstrate that the proposed LPMA shows a clear throughput enhancement over NOMA and OMA schemes.

\section{ Preliminaries}
In this section, we present the basic algebra concepts of a lattice and some discussions about NOMA and its drawback.
\subsection{Fundamentals of lattice}
Denote the complex domain as ${\mathbb C}$. A principle ideal domain (PID) \cite{lattice2}, denoted by $R$, is a commutative ring which satisfies 1) $R\subset {\mathbb C}$; 2) whenever $ab=0$, $a,b\in R$, either $a=0$ or $b=0$; 3) if $aR=\left\{ar:r\in R\right\}$, $a\in R$, then $a$ belongs to the ideal in $R$, where ideal is defined as the subset of $R$ whose elements are closed under addition and multiplication by an arbitrary element in $R$. Commonly, $R$s are: 1) the ring of integers ${\mathbb Z}$; 2) the ring of Gaussian integers ${\mathbb Z}\left[i\right]$, where ${\mathbb Z}\left[i\right]=\left\{a+bi:a,b\in {\mathbb Z}; i=\sqrt{-1}\right\}$; and 3) the ring of Eisenstein integers ${\mathbb Z}\left[\omega \right]$, where ${\mathbb Z}\left[\omega \right]=\left\{a+b\omega :a,b\in{\mathbb Z}; \omega =\frac{-1+i\sqrt{3} }{2}\right\}$.


\begin{definition}
($R$-Lattice) \cite{lattice}. Let $\Lambda$ denote a set of ${\mathbb C}^{n}$ which forms a $R$-module of rank $N$, i.e., $\Lambda$ is closed under addition and multiplication by scalars in the ring $R$.  It is called a $N$-dimensional $R$-lattice. Suppose that there exist $N$ linearly independent vectors $\mathbf{t}_{1} ,...\mathbf{t}_{N} \in \Lambda $ such that $\Lambda =\left\{\sum _{j=1}^{N}r_{j} \mathbf{t}_{j}  ,r_{j} \in R,\forall j\right\}$. The lattice $\Lambda$ satisfies that: if $\mathbf{x}$, $\mathbf{y}\in \Lambda $, then $\mathbf{x}+\mathbf{y}\in \Lambda $; and if $\mathbf{x}\in \Lambda $, then $-\mathbf{x}\in \Lambda $. A subset $\Lambda'\subset\Lambda$, which is a $R$-module, is called a sub-lattice of $\Lambda$. The quotient group $\Lambda/\Lambda' =\{ \lambda  +\Lambda':\lambda \in \Lambda \}$ of $\Lambda $ naturally forms a partition of $\Lambda$.
\end{definition}

\begin{definition}
(Lattice Quantizer) \cite{lattice}. The nearest neighbour lattice codeword in $\Lambda $ associated with $\mathbf{x}\in {\mathbb C}^{n} $  is returned by the following quantizer
\begin{equation}Q_{\Lambda } \left(\mathbf{x}\right)=\lambda \in \Lambda ;\; \left\| \mathbf{x}-\lambda \right\| \le \left\| \mathbf{x}-\lambda '\right\| ,\; \forall \lambda '\in \Lambda ,\end{equation}
where $\left\| \cdot \right\| $ represents the norm operation.
\end{definition}

\begin{definition} (Modulo Operation) \cite{lattice}. The $\mod \Lambda $ operation returns the quantization error w.r.t. $\Lambda $ and is represented as
\begin{equation}
\mathbf{x}\mod\Lambda =\mathbf{x}-Q_{\Lambda } \left(\mathbf{x}\right).
\end{equation}
\end{definition}

\begin{definition}
(Constitution  ${\pi _A}$ \cite{Construction_pi_A}). Let ${\theta _1},...,{\theta _L}$ be a collection of distinct primes from $R$ and ${{\mathbf{G}}_\ell }$ be the generator matrix of a $\left( {n,{k_\ell }} \right)$ linear code ${\mathcal{C}_\ell }$ over the finite field ${\mathbb{F}_{{q_\ell }}}$ with rate ${{{k_\ell }} \mathord{\left/
 {\vphantom {{{k_\ell }} n}} \right.
 \kern-\nulldelimiterspace} n}$, $\forall \ell  \in \left\{ {1,...,L} \right\}$. Construction ${\pi _A}$ is then given by:
 \begin{enumerate}
   \item generating ${\mathcal{C}_\ell }$ from ${\mathcal{C}_\ell } = \left\{ {{\mathbf{v}} = {\mathbf{G}} \otimes {\mathbf{w}}:{\mathbf{w}} \in \mathbb{F}_{{q_\ell }}^{{k_\ell }}} \right\}$,  $\forall \ell  \in \left\{ {1,...,L} \right\}$;
\item constructing ${\Lambda ^*} = \mathcal{W}\left( {{\mathcal{C}_1},...,{\mathcal{C}_L}} \right)$, where
$\mathcal{W}:{\mathbb{F}_{{q_1}}} \times ... \times {\mathbb{F}_{{q_L}}} \to R/\mathop \Pi \limits_{\ell  = 1}^L {\theta _\ell }$
is a ring isomorphism;
\item tiling ${\Lambda ^*}$ to the entire $\mathbb{C}$ to form $\Lambda  = {\Lambda ^*} + \mathop \Pi \limits_{\ell  = 1}^L {\theta _\ell }{R^n}$.
 \end{enumerate}
\end{definition}

\subsection{NOMA and its drawback}
According to \cite{Cover}, the downlink rate region of a power domain based MA (including NOMA) satisfies
\begin{equation}
\begin{gathered}
  {R_1} \le \log (1 + \frac{{P{\alpha _1}|{h_1}{|^2}}}{{P{\alpha _2}|{h_1}{|^2} + 1}}), \hfill \\
  {R_2} \le \log (1 + P{\alpha _2}|{h_2}{|^2}), \hfill \\
\end{gathered}
\end{equation}
where $|{h_i}{|^2},i \in \left\{ {1,2} \right\}$ denotes the user's channel gain and we assume that $|{h_1}{|^2} < |{h_2}{|^2}$; $P$ denotes the total power of composite constellation; and ${\alpha _i}$ denotes the power allocation coefficient where ${\alpha _1}+{\alpha _2}=1$.

When two users experience the similar channel conditions, i.e., $|{h_1}{|^2} \approx |{h_2}{|^2}$, the sum-rate is then calculated as
\begin{equation}
\begin{aligned}
 {R_{Sum}} \le& \log (1 + \frac{{P{\alpha _1}|{h_1}{|^2}}}{{P{\alpha _2}|{h_1}{|^2} + 1}}) + \log (1 + P{\alpha _2}|{h_2}{|^2}) \\
   =& \log (1 + P|{h_1}{|^2}),
\end{aligned}
\end{equation}
which indicates that the sum-rate of two users degrades to the single user capacity.

To avoid the above challenge, the designed scheduler should meet the geometrical separation which pairs the users based on channel power differentiation. This would result in an excessively high complexity when cellular system has a high user density and huge traffic demand.

\section{Detailed design}
\begin{figure*}[!htp]
\centering
\includegraphics[width=6.3in]{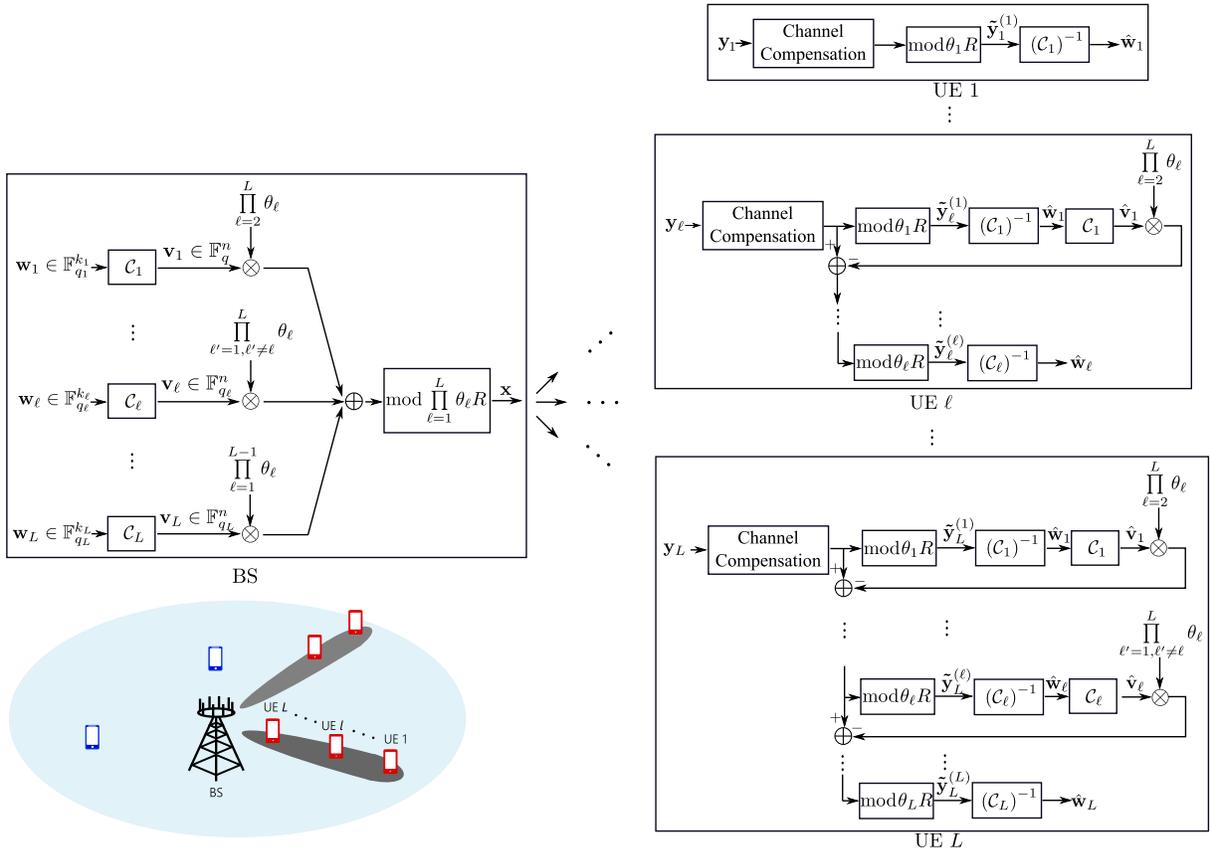}
\caption{Schematic illustration of the proposed LPMA.}
\label{fig_system_model_LPMA}
\end{figure*}
In this section, we provide the detailed design for the proposed LPMA. The systematic illustration of LPMA is shown in Fig. 1 with notations explained in Table 1. The detailed explanations of Fig. 1 are provided in the following subsections.
\begin{table}
\centering
\small
\caption{Notations of Fig. 1}
\begin{tabular}{|c|p{2.3in}|}
\hline
\texttt{Notations} & \texttt{Physical Meanings} \\
\hline
$\mathbf{w}_{\ell } $ & \texttt{UE $\ell $'s desired signal} \\ \hline
${\rm {\mathcal C}}_{\ell } $ & \texttt{The FEC encoder for UE $\ell $} \\ \hline
$\mathbf{v}_{\ell } $ & \texttt{The codeword generated from $\mathbf{w}_{\ell } $} \\ \hline
${\mathbb F}_{q_{\ell } }^{} $ & \texttt{The finite field with size $q_{\ell } $, where $q_{\ell } $ should be a prime} \\ \hline
$k_{\ell } $ & \texttt{The length of $\mathbf{w}_{\ell } $} \\ \hline
$n$ & \texttt{The length of resulting FEC codeword} \\ \hline
$\theta _{\ell } $ & \texttt{The weight coefficient which is the prime and used to form the lattice code level.} \\ \hline
$R$ & \texttt{PID, which can be: \begin{enumerate}
                          \item the ring of integers $\mathbb{Z}$;
                          \item the ring of Gaussian integers $\mathbb{Z}\left[ i \right]$;
                          \item the ring of Eisenstein integers $\mathbb{Z}\left[ \omega  \right]$.
                        \end{enumerate}}\\ \hline
$\left({{\mathcal C}}_{\ell } \right)^{-1} $ & \texttt{The FEC decoder for UE $\ell $} \\
\hline
\end{tabular}
\end{table}

\subsection{Transmissions of lattice superposition coding}
Recall the Section II A, we propose to design the mapping function of superposition coding based on Construction ${\pi _A}$, given by
\begin{equation}
  \mathbf{x} = \beta[\mathcal{W}(\mathbf{v}_1,\ldots,\mathbf{v}_L) + \mathbf{u}],\\
\end{equation}
where ${\mathbf{x}}$ represents the resulting lattice codeword for $L$ user equipments (UE); $\mathbf{u}$ is a fixed dither which minimizes the average transmission power (e.g., the entire constellation has the zero mean); $\beta$ is the scaling factor to meet the power constraint; and $\mathcal{W}$ denotes the mapping function, given by
\begin{equation}
\mathcal{W}\left( {{{\mathbf{v}}_1},...,{{\mathbf{v}}_L}} \right)\triangleq \left[ {\sum\limits_{\ell  = 1}^L {{{\mathbf{v}}_\ell }} \mathop \Pi \limits_{\ell ' = 1,\ell ' \ne \ell }^L {\theta _{\ell '}}} \right]\,\bmod \,{\mkern 1mu} \mathop \Pi \limits_{\ell  = 1}^L {\theta _\ell }R,
\end{equation}
which maps a linear codeword ${{\mathbf{v}}_\ell }$ generated from ${\mathcal{C}_\ell }$ over the finite field ${\mathbb{F}_{{q_\ell }}}$ into a lattice codeword in an element-wise manner. Here is a simple example: let $R$ be $\mathbb{Z}$ and $L=2$, then ${\mathbf{x}} = \beta \cdot\left[ {\mathcal{W}\left( {{{\mathbf{v}}_1},{{\mathbf{v}}_2}} \right) + {\mathbf{u}}} \right] = \beta \left[ {\left( {{{\mathbf{v}}_1}{\theta _2} + {{\mathbf{v}}_2}{\theta _1}} \right)\bmod \,{\mkern 1mu} {\theta _1}{\theta _2} + \mathbf{u}} \right]$, where ${\theta _1}$ and ${\theta _2}$ are primes from $\mathbb{Z}$.

Analogous to the power allocation principle of NOMA, LPMA guarantees the user fairness through the structural power assignment, i.e, the codeword of the user with a poor channel condition is weighted by a relatively large product while the user with a good channel condition is weighted by a relatively small product. A simple example is: suppose that UE 1 is the far user while UE 2 is the near user. Let ${\theta _1}=2$ and ${\theta _2}=7$. Then ${\mathbf{x}} = \beta \left[ {\left( 7{{\mathbf{v}}_1} + 2{{\mathbf{v}}_2} \right)\bmod \,{\mkern 1mu} 14 + \mathbf{u}} \right]$ so that the far user's poor channel condition is compensated by this large prime ${\theta _2}=7$ .

\subsection{Reception of of lattice superposition coding}
Suppose a base station (BS) serves $L$ UEs such that the superimposed signal at the UE $\ell $ can be presented as
\begin{equation}
{{\mathbf{y}}_\ell } = {h_\ell }{\mathbf{x}} + {{\mathbf{z}}_\ell },
\end{equation}
where ${\mathbf{x}} \in \Lambda$ is the transmitted lattice codeword from the BS with dimension $n$ and transmission power $P$; $h_{\ell } $ is the channel coefficient from the BS to UE $\ell $; and ${{\mathbf{z}}_\ell } \sim \mathcal{C}N\left( {0,{\sigma ^2}{{\mathbf{I}}^n}} \right)$ is the noise.

After the channel compensation, the received signal is equivalent to
\begin{equation}
{{\mathbf{\tilde y}}_\ell } = {\raise0.7ex\hbox{${{{\mathbf{y}}_\ell }}$} \!\mathord{\left/
 {\vphantom {{{{\mathbf{y}}_\ell }} {{h_\ell }}}}\right.\kern-\nulldelimiterspace}
\!\lower0.7ex\hbox{${{h_\ell }}$}}-\mathbf{u} =  {\mathbf{x}}-\mathbf{u} + {{\mathbf{z}}_{\ell ,{\text{eqv}}}},
\end{equation}
where  ${{\mathbf{z}}_{\ell ,{\text{eqv}}}} \triangleq {\raise0.7ex\hbox{${{{\mathbf{z}}_\ell }}$} \!\mathord{\left/
 {\vphantom {{{{\mathbf{z}}_\ell }} {{h_\ell }}}}\right.\kern-\nulldelimiterspace}
\!\lower0.7ex\hbox{${{h_\ell }}$}}$ is the equivalent noise with the variance $\sigma _{{\rm eqv,}\ell }^{2} ={\raise0.7ex\hbox{$ \sigma ^{2}  $}\!\mathord{\left/ {\vphantom {\sigma ^{2}  \left|h_{\ell } \right|^{2} }} \right. \kern-\nulldelimiterspace}\!\lower0.7ex\hbox{$ \left|h_{\ell } \right|^{2}  $}} $ and the operation of $-\mathbf{u}$ is to remove the fixed dither. The proposed SIC can exploit the layered structure of the constructed lattice $\Lambda $ and extract the UEs' desired signals from ${{\mathbf{\tilde y}}_\ell }$.

Without loss of generality, we assume that the order of channel conditions of all UEs is: UE $1<{\dots}<$UE $\ell<{\dots}<$UE $L$. As such, the $\ell$-th UE, $\forall \ell  \in \left\{ {2,...,L} \right\}$ can apply the successive decoding from the code level of UE $1$ to that of itself.

Take UE $L$ as an example: as seen in Fig. 1, the first code level can be `peeled off' through the modulo operation w.r.t. ${\theta _1}$, given by
\begin{equation}
{\mathbf{\tilde y}}_L^{(1)} = {{\mathbf{\tilde y}}_L}{\text{ mod }}{\theta _1}R,
\end{equation}
where the interference from other UEs are cancelled out due to the property of co-primes. Feeding ${\mathbf{\tilde y}}_L^{(1)}$ into the decoder of the first code level, namely ${\left( {{\mathcal{C}_1}} \right)^{ - 1}}$, then  the decoded message ${{\mathbf{\hat w}}_1}$ can be extracted, which is then re-encoded into ${{\mathbf{\hat v}}_1}$ for reconstructing the first code level.

The second code level can be obtained from subtracting the reconstructed first code level and taking module operation w.r.t. ${\theta _2}$, given by
\begin{equation}
{\mathbf{\tilde y}}_L^{(2)} = \left[ {{{{\mathbf{\tilde y}}}_L} - {{{{\mathbf{\hat v}}}_1}} \mathop \Pi \limits_{s' = 2}^L {\theta _{s'}}} \right]{\text{mod }}{\theta _2}R.
\end{equation}
Then ${\mathbf{\tilde y}}_L^{(2)}$ is fed into the decoder of the second code level, namely ${\left( {{\mathcal{C}_2}} \right)^{ - 1}}$, so that ${{\mathbf{\hat w}}_2}$ can be decoded, which is then re-encoded into ${{\mathbf{\hat v}}_2}$ for reconstructing the second code level.

In a similar fashion, for the code level $\ell  \in \left\{ {2,...,L{\text{ }}} \right\}$, the decoder subtracts all reconstructed interference from the previous levels via forming
\begin{equation}
{\mathbf{\tilde y}}_L^{(\ell )} = \left[ {{{{\mathbf{\tilde y}}}_L} - \sum\limits_{s = 1}^{\ell  - 1}  {{{{\mathbf{\hat v}}}_s}}  \mathop \Pi \limits_{s' = 1,s' \ne s}^L {\theta _{s'}}} \right]{\text{mod }}{\theta _\ell }R.
\end{equation}
The above decoding method is referred to as the modulo lattice operation with SIC (MLO-SIC).

In some scenarios, the decoding latency is a sensitive issue. To reduce the decoding latency, the modulo lattice operation with parallel interference cancellation (MLO-PIC) is proposed, as seen in Fig. 3. Due to the property of co-prime, the $\ell $-th code level can be `peeled off' via the modulo operations w.r.t. ${\theta _\ell }$,  $\forall \ell  \in \left\{ {1,...,L} \right\}$,
\begin{equation}
\begin{gathered}
  {{{\mathbf{\tilde y}}}_1} = {{\mathbf{y}}_1}{\text{ mod }}{\theta _1}R  \\
   \vdots  \hfill \\
  {{{\mathbf{\tilde y}}}_\ell } = {{\mathbf{y}}_\ell }{\text{ mod }}{\theta _\ell }R  \\
   \vdots  \hfill \\
  {{{\mathbf{\tilde y}}}_L} = {{\mathbf{y}}_L}{\text{ mod }}{\theta _L}R
\end{gathered},
\end{equation}
so that no successive decoding is needed and any user only extracts its desired signals without knowing others'. This property facilitates the massive connectivity in the high SNR regime.

\begin{figure}[!htp]
\centering
\includegraphics[width=2.45in]{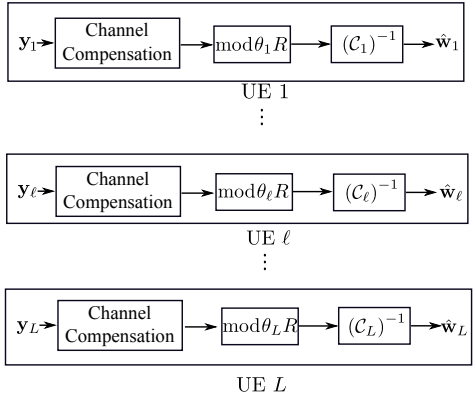}
\caption{MLO-PIC decoding in LPMA.}
\label{fig_system_model_LPMA}
\end{figure}

\begin{figure}[!htp]
\centering
\includegraphics[width=3.5in]{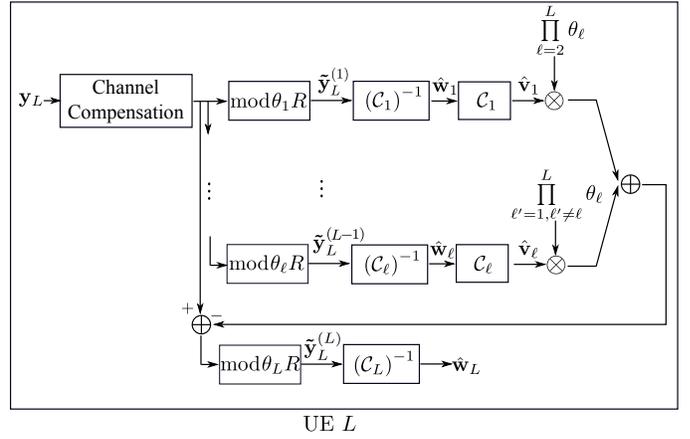}
\caption{Hybrid decoding in LPMA.}
\label{fig_hybrid_decoding}
\end{figure}

Moreover, a hybrid decoding (w.r.t. to the nearest UE $L$, as seen in Fig. \ref{fig_hybrid_decoding}) where MLO-SIC and MLO-PIC are combined, is also proposed, which seeks for the balance between the error performance and decoding latency. For example, in this decoding approach, UE $L$ adopts MLO-PIC to first cancel out the contribution from all the other UEs’ codewords. Then MLO-SIC is used to reconstruct all other UEs' code levels which are then subtracted from the received signal to improve the reliability. It is worth noting that in this hybrid decoding method, PIC-decoded layers and SIC-decoded layers can be flexibly adjusted based on the realistic demands.

\subsection{User scheduling and  pairing}
According to \cite{Construction_pi_A}, the same finite field can be simultaneously isomorphic to more than one distinct lattice partitions. This implies that LPMA can still multiplex users even when they experience the similar channel conditions. An example is given as follows.
\begin{example}
Suppose that UE 1 and UE 2 experience the similar channel conditions. LPMA allows the BS to assign two linear codes $\mathcal{C}_1$ and $\mathcal{C}_2$ over the same field to UE 1 and UE 2. As the same finite field can be simultaneously isomorphic to more than one distinct lattice partitions, two distinct Eisenstein primes, e.g., ${\theta _1} = 2 + 3\omega $ and  ${\theta _2} = 3 + 2\omega$, can be associated with $\mathcal{C}_1$ and $\mathcal{C}_2$. This is because the isomorphisms ${\mathbb{F}_7} \cong \mathbb{Z}[\omega]/{\theta _1}\mathbb{Z}[\omega]$ and ${\mathbb{F}_7} \cong \mathbb{Z}[\omega]/{\theta _2}\mathbb{Z}[\omega]$ hold. The resulting codeword $\mathcal{C}_1$ is weighted by ${\theta _2}$ while $\mathcal{C}_2$ is weighted by ${\theta _1}$. As $\theta_1$ and $\theta_2$ are relatively prime, UE 1 can adopt the $\bmod {\theta _1}\mathbb{Z}[\omega]$ operation to null out UE 2's contribution while UE 2 nulls out UE 1's contribution by the $\bmod {\theta _2}\mathbb{Z}[\omega]$ operation.
\end{example}

The above \emph{Example 1} indicates that due to the property of co-prime and modulo lattice operation, both users still can be separated even though they are allocated with an equal power (due to the same finite field). In the following example, we can observe how LPMA reduces the complexity of scheduling.
\begin{figure}[!htp]
\centering
\includegraphics[width=3.2in]{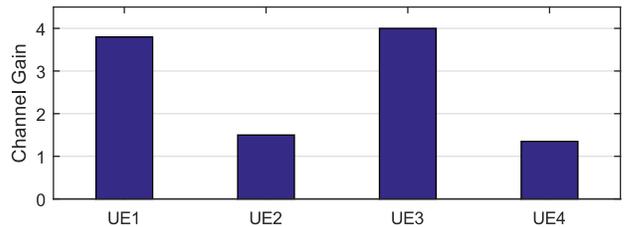}
\caption{Channel gains of different users (Example 2).}
\label{fig_channel_gain}
\end{figure}
\begin{example}
Suppose there are 4 UEs randomly dropped in the cellular with the channel gains given in Fig. \ref{fig_channel_gain}. They are paired into 2 groups for downlink transmissions where 2 orthogonal resources are correspondingly allocated to those 2 groups. NOMA/LPMA is used to multiplex users in the same group. According to \emph{Example 1}, LPMA can randomly pair the users by using the round-robin method as it is robust to scenario where the channel gains of users are equally strong. However, we note that NOMA needs to pair the users as \{UE 1, UE 2\}, \{UE 3, UE 4\} or \{UE 1, UE 4\}, \{UE 3, UE 2\}. If the random pairing is applied in NOMA, there will be $33.3\%$ probability of capacity degradation.
\end{example}

\subsection{Modulation, coding scheme, and power allocation}
We note that after taking the modulo lattice operation, the contributions of other users' signals are cancelled out while the noise distribution would be changed, i.e., the noise is folded into the modulo operation region. This effect is negligible when the noise variance is small. This implies that the MCS of LPMA can directly adopt that of OMA so that the compatibility with LTE system is guaranteed and the complexity is low.

Based on the feedback SINR of UE, the BS can check a look up table, which stores the throughput v.s. SINR to get the corresponding throughput. The returned throughput determines this UE's distinct prime. Once all UEs' primes are determined, the structural power assignment of LPMA (namely, assigning the product of primes to each user) is determined accordingly. Because of the structural property of lattice, the power adaption is not required in LPMA.

\section{ Performance Evaluation}
The performance evaluation is provided in this section, where Fig. 6 shows the comparison of user throughput among OMA, NOMA, and the proposed LPMA. The simulation parameters are given in the following table.
Based on Fig. 6, we can observe that: 1) compared with NOMA, the proposed LPMA enhances the user throughput by $28.6\%$. 2) compared with OMA, our proposed LPMA enhances the user throughput by $63.7\%$.

\begin{table}[!htp]
\centering
\small
\caption{The parameters of simulation in Fig. 4}
\begin{tabular}{|c|p{1.6in}|}
\hline
\texttt{Cellular Layout}:  & \texttt{Hexagonal, wrapped around} \\ \hline
\texttt{Topology} & \texttt{$7$ sites (full buffer, no sectorization)} \\ \hline
\texttt{Bandwidth} & \texttt{$10$Mhz} \\ \hline
\texttt{Tx Antenna No.} & \texttt{$1$ (omni-directional)} \\ \hline
\texttt{Rx Antenna No.} & \texttt{$1$ (omni-directional)} \\ \hline
\texttt{No. of UEs per cell} & \texttt{$2$} \\ \hline
\texttt{BS-to-BS distance} & \texttt{$500$m} \\ \hline
\texttt{Channels} & \texttt{i.i.d Rayleigh fading} \\ \hline
\texttt{BS Tx power} & \texttt{$46$dBm} \\ \hline
\texttt{Thermal noise density} & \texttt{$-174$dBm/Hz} \\ \hline
\texttt{Rx Noise figure} & \texttt{$5$dB} \\ \hline
\texttt{Path loss model} & \texttt{$21.5 + 36.7\log_{10}(D)$, $D$ in km} \\ \hline
\end{tabular}
\end{table}

\begin{figure}[!htp]
\centering
\includegraphics[width=3.3in]{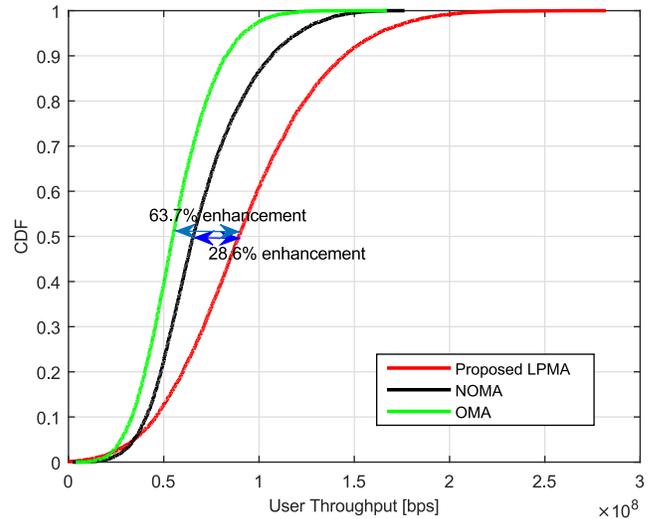}
\caption{Throughput comparison among the proposed LPMA, NOMA and OMA.}
\label{fig_system_model_LPMA}
\end{figure}
\section{Conclusive Remarks and future work}
In this paper, we proposed a non-orthogonal downlink multiuser transmission scheme, termed lattice partition multiple access (LPMA), which is robust to channel aging, and enjoys low processing overhead. A lattice code based superposition coding is chosen as the basis for the new multiple access design. The BS encodes the multiuser signals into different lattice code levels and each UE extracts their desired signal by performing a modulo lattice operation and SIC/PIC (MLO-SIC/MLO-PIC) decoding. The proposed LPMA exploits the interference by utilizing the structural property of lattice codes. Unlike NOMA, LPMA can still multiplex users when they experience the equally strong channels. This advantage makes it possible to design a low complexity scheduling algorithm. Moreover, LPMA offers a simple MCS design and a structural power allocation. We demonstrated that the proposed LPMA shows a clear throughput enhancement over the current NOMA and OMA schemes.


\end{document}